\documentclass[aip,reprint,letterpaper,numerical]{revtex4-1}
\usepackage{color}
\usepackage{geometry}
\usepackage{graphicx}
\usepackage{amssymb}
\usepackage{bm}
\usepackage{amsmath}

\newcommand{\mat}[1]{\mathrm{\underline{\underline{#1}}}}

\begin{document}

\title{Are hydrodynamic interactions important in the kinetics of
  hydrophobic collapse?}
 
 \author{Jingyuan Li}
 \affiliation{Department of Chemistry,
 Columbia University, 3000 Broadway, MC 3103, New York, NY 10027}
 \affiliation{Chinese Academy of Sciences Key Lab
  for Biomedical Effects of Nanomaterials and Nanosafety, Institute of High Energy
  Physics, Chinese Academy of Sciences, Beijing 100049, China}
  
 \author{Joseph A. Morrone}
 \affiliation{Department of Chemistry,
  Columbia University, 3000 Broadway, MC 3103, New York, NY 10027}

\author{B. J. Berne} 
\email{bb8@columbia.edu}
 \affiliation{Department of Chemistry, Columbia University, 3000
  Broadway, MC 3103, New York, NY 10027}

\begin{abstract}
  We study the kinetics of assembly of two plates of varying
  hydrophobicity, including cases where drying occurs and water
  strongly solvates the plate surfaces.  The potential of mean force
  and molecular-scale hydrodynamics are computed from molecular
  dynamics simulations in explicit solvent as a function of particle
  separation.  In agreement with our recent work on nanospheres
  [J. Phys. Chem. B \textbf{116,} 378 (2012)]
  regions of high friction are found to be engendered by large and
  slow solvent fluctuations. These slow fluctuations can be due to either
  drying or confinement. The mean first passage times for assembly are
  computed by means of molecular dynamics simulations in explicit
  solvent and by Brownian dynamics simulations along the reaction
  path.  Brownian dynamics makes use of the potential of mean force
  and hydrodynamic profile that we determined.  Surprisingly, we find
  reasonable agreement between full scale molecular dynamics and
  Brownian dynamics, despite the role of slow solvent relaxation in
  the assembly process. We found that molecular scale hydrodynamic
  interactions are essential in describing the kinetics of assembly.
\end{abstract}

\maketitle

\section{Introduction}
\label{sec:Introduction} Hydrophobic interactions are important in
structural biology and nanoscience where they are often the major
driving force in self-assembly.~\cite{Berne:2009dg,
WALLQVIST:1995uw, chandler:2005p75, Lum:1999p624, Jamadagni:2010bc,
Margulis:2003wo, willard08, patel10, Zangi:2011p581} For
example, hydrophobic interactions are acknowledged to play a major
role in the formation and folding of proteins.~\cite{ruhong04, liu05}
A classic example of hydrophobic assembly is when two plates that
bear little attraction to water are separated by less than some
critical separation, capillary evaporation occurs in the inter-plate
region, thereby driving association.  

Although a great deal of work has concentrated on the structural and
thermodynamic aspects of hydrophobic assembly, less is known about
the kinetic mechanisms involved in the association process. Prior studies have investigated the rate of  evaporation of solvent confined between two hydrophobic surfaces,~\cite{bolhuis2000,leung2003,Margulis:2003wo,luzar2004,sharma12} although the full association pathway remains largely unexplored.  One possible route to modeling the kinetics is within the framework of
Brownian (Smoluchowski) dynamics, a coarse grained description in
which the solvent degrees of freedom are not explicitly treated and
where the dynamics of the heavy bodies are instead modeled in a
stochastic bath.  The effect of solvent is encoded in the potential
of mean force (PMF) between bodies and the friction coefficient.  If
hydrodynamic interactions (HI) are included,  then the friction
coefficients depend on the positions of the heavy bodies. Within this
framework, one can simulate protein diffusion and association using the
appropriate equations together with the potential of mean force and
the (position dependent) friction tensor as input.~\cite{mccammon78}
For example, such an approach has been utilized to study crowded
cellular environments, where the diffusion of proteins is thought to
slow down considerably due in part to hydrodynamic
interactions.~\cite{Ando:2010p99}

Most typically, hydrodynamic interactions that are utilized in
Brownian dynamics simulation are computed within a continuum
description of the solvent.~\cite{Brady:1988p970, Brady:1988tz}  This
description, however breaks down for inter-particle length scales of
1-2 nm.~\cite{Thomas:2009tj, Bocquet:cy} On such length scales,
fluctuations which accompany solvent layering or capillary drying
could have an enormous effect on hydrodynamic interactions. In
recent work, we have shown how the nature of solvent confined
between nanoscopic spheres determines the role of HI
in assembly.~\cite{joe12}  There is a growing interest in
the behavior of water molecules in confined
geometries.~\cite{rossky07,Hummer:AR08,Baron:JACS10}  The diffusion
constant and the hydrogen bond lifetime of water molecules in
confined environments are distinct from those in the bulk.~\cite{Li05} 
Furthermore, similar signatures have been recently
observed in simulations of crowded protein
solutions.~\cite{harada12}  We believe such effects can potentially
play an important role in the transport and association of
nanoscopic bodies.

In this paper, we extend our previous study of nanospheres to the
study of hydrodynamic interactions and association kinetics of
plates of varying hydrophobicity.  We calculate the 
friction on two parallel plates as a function of their
separation from molecular dynamic simulations with explicit water
using a technique previously employed for spherical
solutes.~\cite{Bocquet:1997p103,joe12} We also calculate the
potential of mean force between the two plates as a function of
their separation. Both the molecular-scale effects of hydrodynamic
interactions and the free energy profile are found to be highly
correlated with the behavior of water molecules in the inter-plate
region, in agreement with our prior work.~\cite{joe12}

We presently consider three types of graphene-like plates: plates
with a ``fully'' attractive solute-solvent interaction potential,
plates with a ``reduced'' attractive solute-solvent potential, and
plates with a purely repulsive solute-solvent potential.  For the
plates with either a reduced attractive potential or a purely
repulsive potential, capillary evaporation occurs when the inter-plate
separation is smaller than a critical value. In contrast, capillary
evaporation does not occur for plates with a fully attractive
solute-solvent potential. As the fully attractive plates approach each
other, we observe solvent layering with water molecules being
eventually expelled from the inter-plate region due to steric
repulsion. Both dewetting in the former cases and solvent layering in
the latter case greatly affect the solvent fluctuations in the
inter-plate region and give rise to molecular scale effects in the
hydrodynamic interactions. We compute the spatial dependence of the
friction tensor and investigate the relation between the static and
dynamic fluctuations of water density in the inter-plate region and
this property.

Given the potential of mean force and position dependent friction
coefficient we compute the rate of diffusion controlled association
by means of Brownian dynamics (BD) simulations. The predictions of
BD are tested against the results garnered from a set of molecular
dynamics (MD) simulations of assembly in explicit solvent.  This serves
as a test of the validity of the Markovian approximation inherent in
Brownian dynamics simulation, and the ability of coarse-grained
stochastic dynamics to adequately capture the kinetic effects
associated with the bath.  We find that there is 
reasonable agreement between the two simulations when molecular
scale hydrodynamic interactions are included in the BD, but marked
disagreement when hydrodynamic interactions are neglected. The observation
that BD with hydrodynamic interactions is in agreement with the MD result is
somewhat surprising since, as we show, slow solvent fluctuations
play an important role in the assembly process, and we would thus
expect non-Markovian effects to be of importance. The agreement
between BD and MD indicates that the non-Markovian effects may be
encoded in the spatial dependence of the friction coefficient in an
average way.

\section{System and Method}
\label{sec:Method} We modeled the hydrophobic plate as a
graphene-like sheet with an area of $(1.2\times1.3)$ nm$^2$ as shown
in Fig. 1. The plate size was chosen so as to facilitate both nanometer scale
drying transitions and comparison with our prior work on nanospheres.
In order to vary the hydrophobicity we studied three
types of plates, all with a carbon-carbon bond length of $0.145$ nm,
and a Lennard-Jones diameter of $\sigma_{CC} = 0.35$ nm but with
different interaction potential well depths: (a) full Lennard-Jones
(LJ) interaction with $\epsilon_{CC}$ = 0.276
kJ/mol,~\cite{werder03} (b) reduced LJ interaction with
$\epsilon_{CC}$ = 0.055 kJ/mol, and (c) purely repulsive
Weeks-Chandler-Anderson (WCA) truncation of the full LJ potential
~\cite{Weeks:1971us}. For convenience, these three types of plates
are denoted as (full) LJ plates, reduced LJ plates and WCA plates in
the present work.  Solute-solvent interactions are given by the geometric mean
of the respective water and solute parameters. 

In order to calculate the relative friction coefficient, two
parallel plates are placed perpendicular to the z-axis and a series
of simulations are run with the plates fixed at various separations, and all internal
degrees of freedom are frozen.
The inter-plate separation ranges from 0.4 to 2 nm (for the full and
reduced LJ plates) or to 2.2 nm (for the WCA plates).  The solvent-induced mean force is computed from such fixed configurations, although a finer grid spacing in $z$ is employed.  In order to obtain the potential of mean force (PMF), this quantity is then integrated in combination with the direct plate-plate interactions.

In the simulations of the association process, two plates are initially
placed  $\approx 1.8$ nm apart. The lower plate is
fixed to its position and a biased force along  the
$-\,z$ direction is applied to
the upper plate. Harmonic potentials are utilized to treat the intramolecular stretches ($k_b= 392721.8$ kJ/mol nm$^{-2}$ and $d_0=0.14$ nm) and bends ($k_\theta = 5271.84$ kJ/mol rad$^{-2}$ and $\theta=120^\circ$).  Snapshots from the association of reduced LJ plates
and representative trajectories plotted along the reaction
coordinate are shown in Fig. 1.  The results from these two sets
of calculations are discussed in Section \ref{sec:fricresults} and
Section \ref{sec:association}, respectively.  In addition, we determine the
friction coefficient on a single plate, in a system containing only one
plate, by pulling it at a constant speed perpendicular to its plane
and computing the drag force along its normal direction.

All systems are solvated by TIP4P water.~\cite{tip4p}  The size of the
solvation box of the full LJ and reduced LJ system is
$4\times4\times4$ nm$^3$, and for the WCA system the box size is
increased to $5\times5\times6$ nm$^3$.  This is done in order to
accommodate the large volume fluctuation that arising from the strong
dewetting transition.

In the calculation of the friction, each system is equilibrated with
a 2 ns NPT simulation (P= 1 atm, T = 300 K). The Berendsen method
~\cite{berendsen} and the stochastic velocity rescaling
method~\cite{Bussi:2007p114} are chosen for the barostat and
thermostat, respectively. For each system, there are 12-18 NVE
production runs of 4 ns to compute the friction coefficient. In
order to facilitate energy conservation, double precision routines
are utilized for all NVE runs. To study the dragging force of the
single plate, we choose four pulling rates ranging from 0.5 to 3
nm/ns. There are at least 12 NPT runs for each pulling rate that are
utilized to obtain an accurate estimate of dragging force.  We
perform at least 55 runs with various initial configurations for
each system to fully characterize the association process of two
plates and estimate the distribution of mean passage times. All
simulations are performed using GROMACS 4.5.3 and the Particle-Mesh Ewald technique is utilized 
to treat long range electrostatics.~\cite{pme,gromacs4}

\begin{figure*}
   \begin{center}
     \includegraphics[scale=0.55]{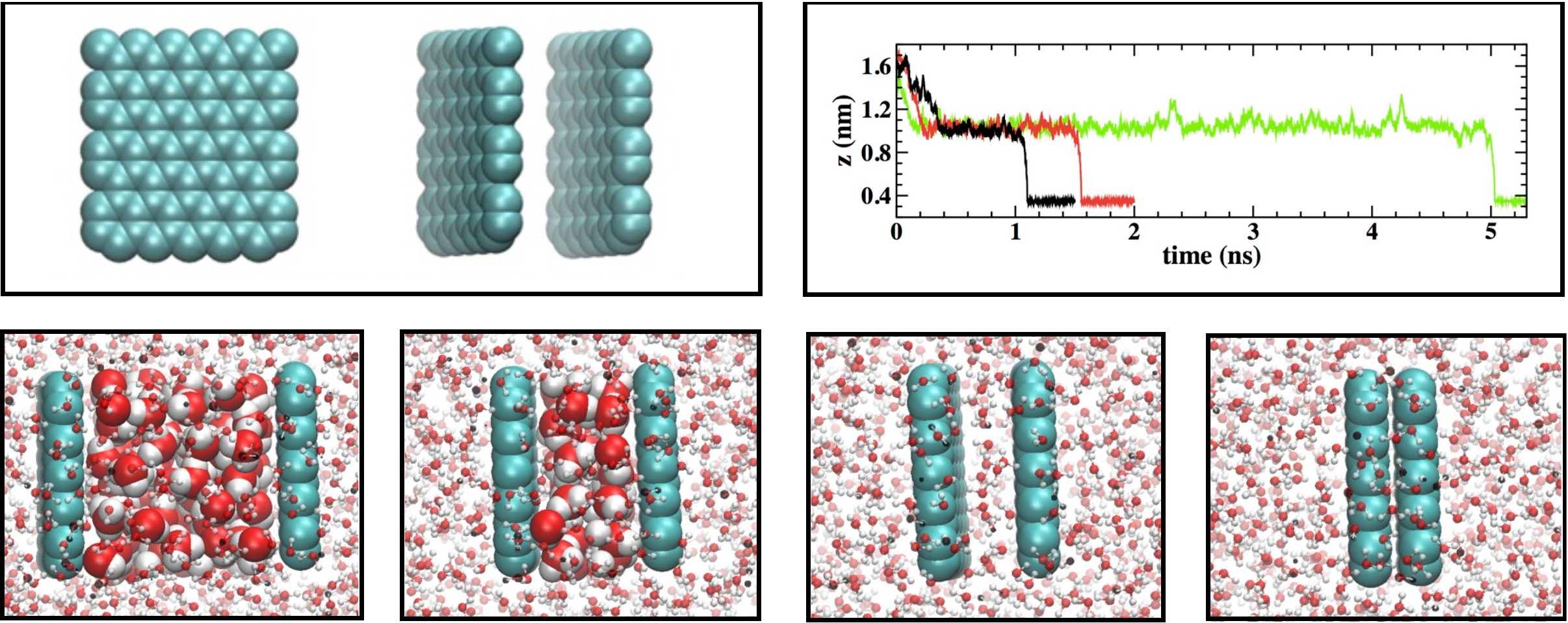}
     \caption{ (Upper left panel) The top view and side view of two parallel graphene-like plates. (Upper right panel)  Representative association trajectories of two
       plates with the ``reduced'' LJ interaction.  (Lower panels)
        Configurations taken from the association of the reduced LJ plates, corresponding to an inter-plate separation of (from left to right): $z$ = 1.7 nm
       (initial configuration), 1.0 nm (where the association stagnates), 0.7 nm (during the driving induced collapse) and 0.35 nm (when the plates come into
       contact). The plates and water molecules in the inter-plate region are depicted by a space filling representation, and the snapshots are rendered with VMD.~\cite{vmd96}}
    \end{center}
  \label{fig1}
\end{figure*}

\section{Brownian dynamics with hydrodynamic interactions}
\label{sec:HI}

Within the framework of Brownian dynamics, hydrodynamic interactions
are encoded in the frictional force. The stochastic equation for the
association process contains contributions from the mean
force, the frictional force, and the Gaussian random force. The
potential of mean force, $W(z)$, includes the contributions from the
direct plate-plate interaction and the solvent-induced interaction.
The separation between two plates, $z$, is treated as the reaction
coordinate.  Hydrodynamic interactions give rise to the spatial
dependence of the relative friction coefficient, $\zeta(z)$ and
through this the spatially dependent diffusion coefficient
$D(z)=kT/\zeta(z)$.

The stochastic BD equation for plate association can be therefore
written as:~\cite{tough86}
\begin{equation}
\dot{z}= -\beta D(z)\frac{\partial}{\partial z} W(z) + \frac{\partial}{\partial z} D(z) + R(z,t) ,
\label{eq:BD}
\end{equation}
where $\beta\equiv 1/k_BT$ and the diffusion coefficient is related to the
``random force'' $R(z,t)$ through the fluctuation dissipation theorem,
\begin{equation}
\langle R(t')R(t)\rangle=2D(z)\delta(t-t').
\label{eq:fluctheorem}
\end{equation}
This equation of motion is an accurate description of the dynamics
in the high friction limit for timescales
greater than the momentum relaxation time, and in which the solvent
time-scale is fast compared to the time scale for the motion of the
heavy bodies. The
  dynamics can alternatively be expressed as a Smoluchowski equation
  for the probability distribution $P(z,t)$ of finding the
  particle at $z$ at time $t$:
\begin{equation}
  \frac{\partial P(z,t)}{\partial t}=\frac{\partial}{\partial z}D(z)\left(\frac{\partial}{\partial z} +\beta W'(z)\right)P(z,t).
\label{eq:smoleq}
\end{equation}
This equation or BD can be used to determine mean first passage times for
diffusion controlled reactions once $W(z)$ and $\zeta(z)$ are known.

In order to apply BD we must determine the PMF and spatial
dependence of the friction coefficient along the reaction path.  As
already pointed out, the continuum treatment of the hydrodynamic
interaction breaks down at small separations.~\cite{Thomas:2009tj,
Bocquet:cy,joe12} In this work, we calculate the friction
coefficient on the molecular length scale from explicit molecular
dynamics simulations.  The friction coefficient can then be
determined from the two-body friction tensor $\zeta_{ij}$. In the
calculation, the plates are fixed as required in the Brownian limit.
The pair friction tensor can be expressed as a time integral of the
correlation functions of the fluctuating force
$\delta\vec{F}_i=\vec{F}_i-\langle\vec{F}_i\rangle$,
\begin{equation}
  \zeta_{ij} = \beta\int^\infty_0 \mathrm{d}t \lim_{n \to \infty}\langle\delta\vec{F}_i(t)\delta\vec{F}_j(0)\rangle,
\label{eq:frictioncoef}
\end{equation}
where by symmetry $\zeta_{12}=\zeta_{21}$ and
$\zeta_{11}=\zeta_{22}.$ Here we focus on the direction parallel to
the inter-plate separation and therefore study the force components
along the plate normal vector.  Eqn. \ref{eq:frictioncoef} is only valid
in the limit where the number of solvent molecules, $n$, approaches
infinity.~\cite{Bocquet:1997p103} In the present case of finite
systems, it is still possible to relate the force autocorrelation
function to the friction tensor, following the procedure of Bocquet
et al.~\cite{Bocquet:1997p103} The friction along the relative
coordinate is given by $(\zeta_{11}-\zeta_{12})/2$ and
depicted in Section \ref{sec:fricresults}.

\section{Results}

  \subsection{ Friction and diffusion coefficients of a single plate }
\label{sec:singleplate}

The friction coefficient perpendicular to the face of a single plate
immersed in solvent may be extracted from the force autocorrelation
function by means of techniques outlined in Ref.
\onlinecite{Bocquet:1994p317}. The resultant calculation yields values
of $1.63\times10^{-11}$, $1.33\times10^{-11}$, $1.0\times10^{-11}$
kg/s, when the solvent-solute interaction is the
full LJ, Reduced LJ, and the WCA potential, respectively.  In
agreement with prior work,~\cite{joe12} the friction experienced by
the body decreases with increasing surface hydrophobicity.  Each
calculation is performed in a periodic box of dimensions given in
Section \ref{sec:Method}.

The friction coefficient may also be extracted from a set of
non-equilibrium simulations where the plate is dragged through the
solvent at a given velocity. As depicted in Fig. 2, the dragging force
at four velocities is determined from MD simulations in explicit
solvent, and grows with increasing velocity. The resultant plot of
dragging force versus pulling rate exhibits a linear relationship with
the slope equal to the friction coefficient.  The extracted values of
the friction are given in Table 1, alongside the results computed from
the force autocorrelation function. The two estimates of the friction
are in close agreement.  The consistency between the results garnered
from two different techniques serves to validate the calculation.

Next, we compare these results to the predictions of continuum
hydrodynamics.  The friction coefficient for a rigid body described by
a set of spherical sites can be evaluated by means of the
Rotne-Prager tensor.~\cite{rotne69}  The translational friction
coefficient may be extracted from the full $3N \times 3N$ mobility
tensor, where $N$ is the number of sites.~\cite{carrasco99}  In the
present calculation, periodic boundary conditions must be taken into
account and are known to have a significant impact on hydrodynamic properties.~\cite{Hasimoto:1959p589,Beenakker:1986tz,Yeh:2004p287} Periodic effects may  be treated by means of replacing the
standard Rotne-Prager tensor with a form that accounts for periodicity
via an Ewald summation.~\cite{Beenakker:1986tz} This approach is
discussed further in the Appendix.

As discussed above, different solvation boxes are utilized in the WCA
and Lennard-Jones systems, and the continuum calculation must be
undertaken for both sizes.  The values of the friction coefficient with stick boundary conditions for
the two box sizes are given
in Table \ref{tab:singlefric}.  The full LJ value is somewhat larger than the continuum result whereas, in line with expectations,
the increasingly hydrophobic plates begin to show larger deviations from the stick boundary condition.

\begin{figure}
   \begin{center}
     \includegraphics[scale=0.27]{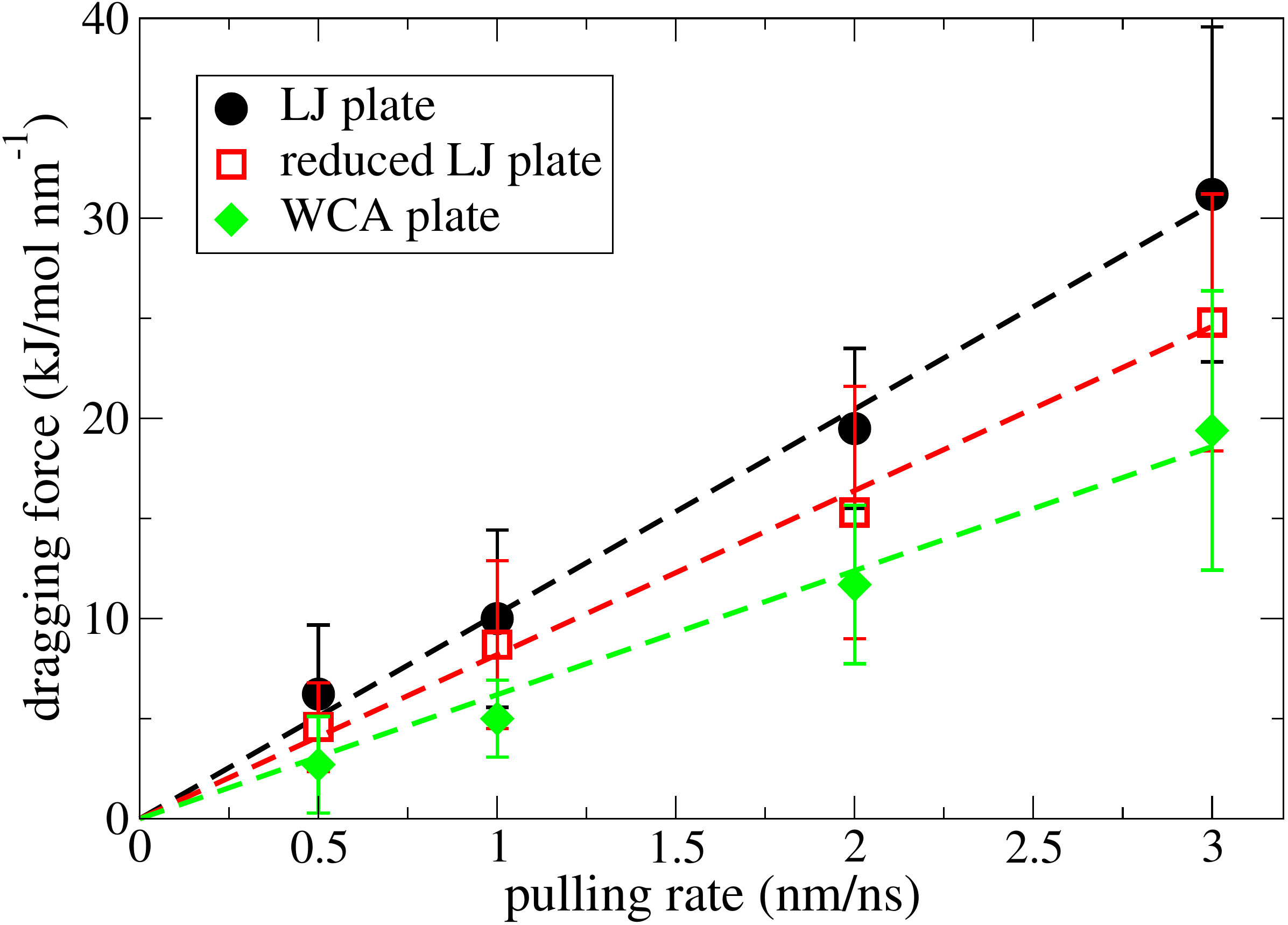}
  \caption{The dragging force of single plate moving at constant velocity. The results for the plate with full Lennard-Jones(LJ) interaction (LJ plate), the plate with reduced LJ interaction (reduced LJ plate), purely
  repulsive Weeks-Chandler-Anderson truncation of the full LJ potential (WCA plate) are depicted in black, red, and green, respectively. The dragging forces increase linearly as the pulling rate increases, and the fitted slopes are equal to the friction coefficient. }
    \end{center}
  \label{fig2}
\end{figure}

\begin{table*}[t]
\begin{center}
\begin{tabular}{|c|c|c|c|c|}
\hline
interaction & box (nm$^3$) & $\zeta_\mathrm{FACF}$ ( $10^{-11}$ kg/s) & $\zeta_\mathrm{PULL}$  ( $10^{-11}$ kg/s)  & $\zeta_\mathrm{STICK}$ ($10^{-11}$ kg/s) \\ \hline
WCA & $5 \times 5 \times 6$ & $1.00 $ & 1.03 & 1.27 \\ \hline
REDUCED & $4 \times 4 \times 4$ & 1.33 & 1.36 & 1.53 \\ \hline
FULL & $4 \times 4 \times 4$ & 1.63 & 1.69 & 1.53 \\ \hline
\end{tabular}
\end{center}
\caption{The friction on a single plate in the $z$-direction, as estimated from the force autocorrelation function, pulling simulations, and continuum hydrodynamics.}
\label{tab:singlefric}
\end{table*}

\subsection{Hydrodynamic and thermodynamic profiles}
\label{sec:fricresults}
Friction coefficients, $\zeta(z)$, and the potential of mean force,
$W(z)$, are depicted as a function of inter-plate separation, $z$, in
Figs. 3, 4, and 5 (for the full LJ, reduced LJ, and WCA plate,
respectively).  The frictional profiles exhibit non-monotonic behavior
as the two plates approach each other. The spatially dependent
features of the molecular scale hydrodynamic interactions display
similar trends to those found in the free energy profile.  This
finding is in agreement with our previous work,~\cite{joe12} and the
recent work of Mittal and Hummer.~\cite{mittal12} In the case of the
full LJ plate, the friction peaks and the free energy increases as a
layer of water is ``squeezed out''~\cite{Zangi:2011p581} at $z=0.88$
nm. The friction coefficient subsequently decreases at the minimum of
the PMF.  The final solvent layer is expelled as the separation
decreases to 0.62 nm, and both the free energy and the friction
coefficient increase. It is important to note that the friction
profile peaks at 0.88 and 0.62 nm which are at the same positions as
the PMF barriers. The peak heights at these two separations are $23.3
(4.2) \times10^{-10}$ and $4.0 (0.4)\times10^{-10}$ kg/s,
respectively.  Standard errors of the mean value are given in
parenthesis.

\begin{figure}
   \begin{center}
     \includegraphics[scale=0.27]{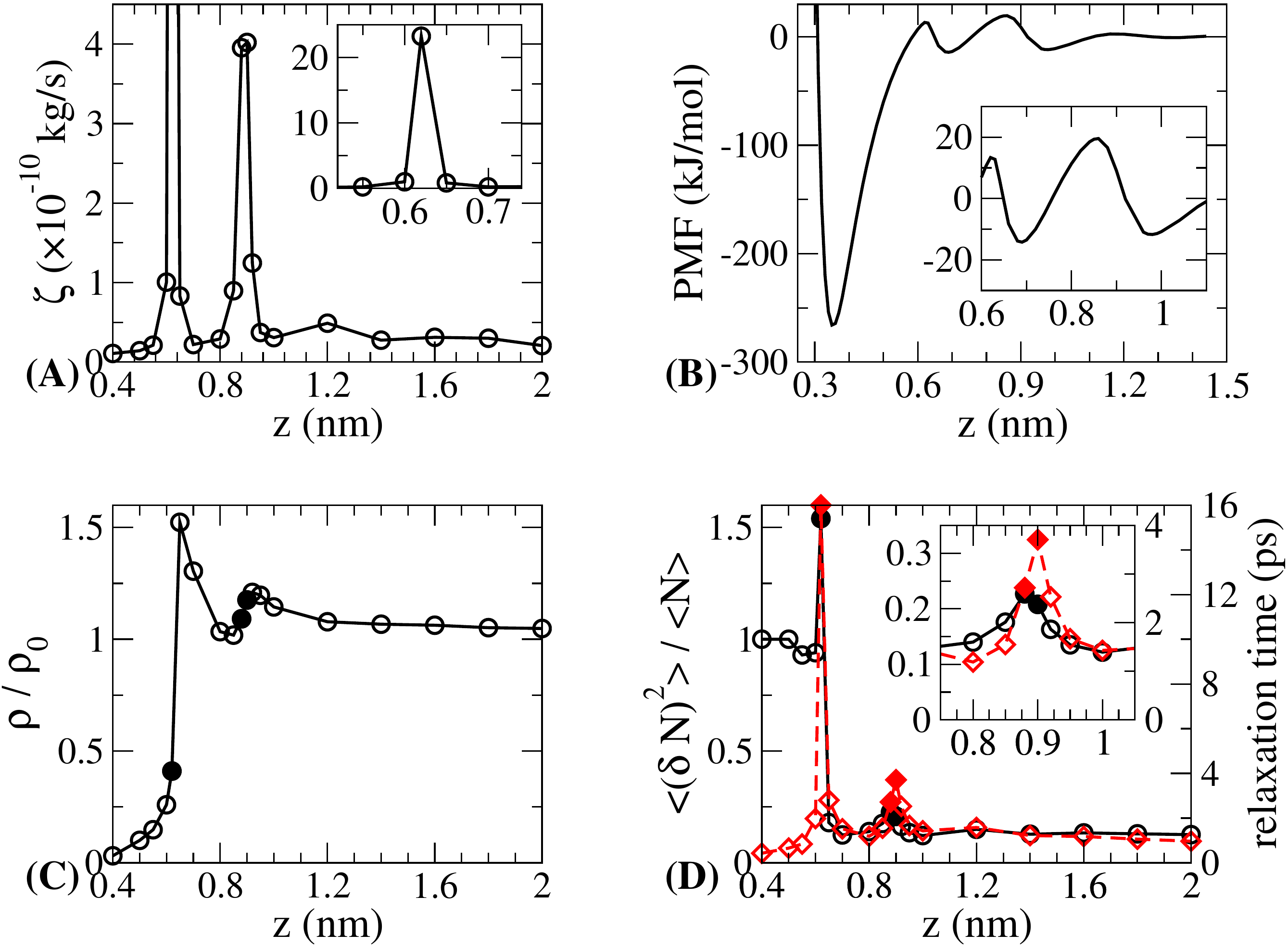}
     \caption{Spatial dependence of relative friction coefficient (A) and the potential of mean force (B) for the association of two LJ plates.
     The friction peaks at $z$ = 0.62 nm and two free energy wells at $z$ = 0.68 and 0.97 nm are depicted in the insets of (A) and (B), respectively.
     (C) The relative density of water in the inter-plate region. (D) The ratio of the variance to the average of the number of water molecules (black)
     and the solvent relaxation time in the inter-plate region (red). For both curves in (D), the peaks at $z \approx$ 0.9 nm are depicted in the inset.
     The values corresponding to the friction peaks are shown as filled symbols in panels (C) and (D).}
    \end{center}
  \label{fig3}
\end{figure}

For the reduced LJ plates, there is a low barrier at $z=0.9$ nm in
the PMF, whereas the WCA plates exhibit barrierless assembly along
the chosen reaction coordinate. In these cases, the corresponding
friction profiles also display non-monotonic behavior. The friction
profile peaks at $z_c$ = 0.92 nm and $z_c$ = 1.35 nm for the reduced
LJ and WCA plates, respectively.  As discussed below, these
distances are in the region of the critical separation for
dewetting.  The corresponding peak heights of the friction
coefficients are $4.1 (0.85) \times10^{-10}$ (reduced LJ) and $2.2
(0.61)\times10^{-10}$ kg/s (WCA). The somewhat large error bar
results from the sizable force fluctuations that are present in the
vicinity of $z_c$.

The friction profiles converge to $2.1\times10^{-11}$ (LJ),
$2.2\times10^{-11}$ (reduced LJ) and $1.6\times10^{-11}$ (WCA) kg/s
at large separations. The standard error in the mean for these values
is $\approx 10\%$.  These numbers are somewhat larger
than expected upon comparison with the results given in Table
1.  This difference may largely result from the finite box size and
the impact of neighboring periodic images. However, the effect of
periodic boundary conditions should not greatly impact the observed
spatial dependence of the short-range hydrodynamics interactions.

In order to better understand the nature of molecular-scale
hydrodynamic interactions, we analyze the density and the static and
dynamic fluctuations in number of water molecules in the inter-plate
region. The static fluctuations are measured by the
ratio of the number variance to the average
$\langle(\delta N)^2\rangle/\langle N\rangle$. An estimate of the
solvent relaxation time can be computed from the integral of the
normalized autocorrelation function of these fluctuations,
$\langle\delta N(t) \delta N(0)\rangle/\langle(\delta N)^2\rangle$.
The resultant plots of these quantities as a function of inter-plate
separation are depicted in Figs. 3-5. Additionally, the values of
the solvent density and fluctuations at the separation that
corresponds to the maximum friction are marked on the curves. For
the LJ plates, the solvent density decreases when the plate
separation decreases from $0.9$ to $0.85$ nm. The ratio of the
variance to the average of water density peaks at $z = 0.88$ nm, and
the relaxation time peaks at $z = 0.90$ nm. Both peaks occur at
separations close to where the friction coefficient peaks. Moreover,
when the separation decreases from $0.65$ to $0.62$ nm the solvent
density sharply decreases. The static fluctuations and relaxation
time grow in the process of water expulsion. Both the static
variance of water density and the relaxation time peak at the same
separation where the friction coefficient peaks ($z_c$ = 0.62 nm).

\begin{figure}
   \begin{center}
     \includegraphics[scale=0.27]{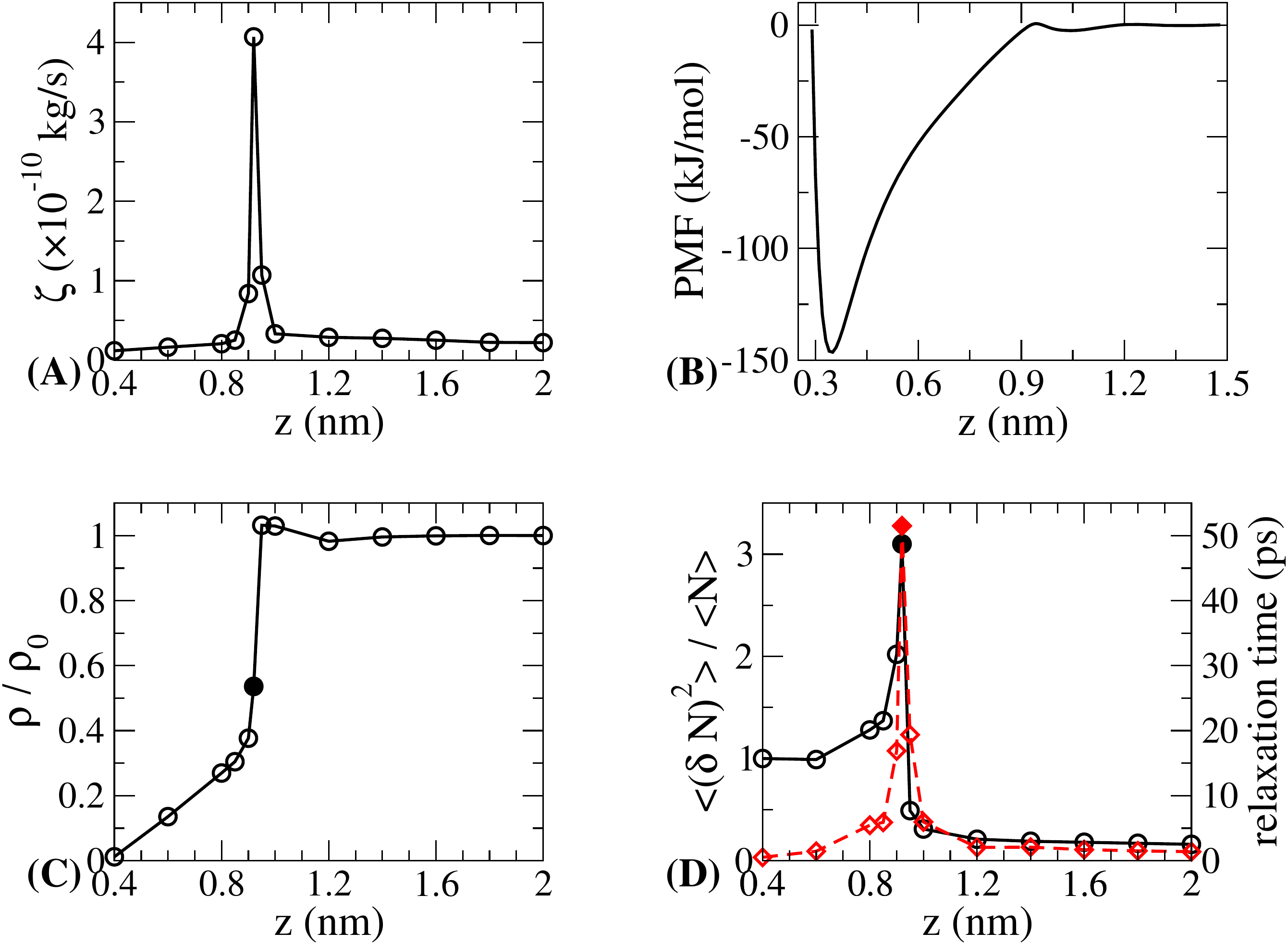}
     \caption{Spatial dependence of relative friction coefficient (A) and
       the potential of mean force (B) for the association of two reduced LJ plates. (C) The
       relative density of water in the inter-plate region. (D) The ratio of the variance to the average of the number of water molecules (black)
       and the solvent relaxation time in the inter-plate region (red). The value corresponding to the friction peak is shown as filled symbols in panels (C) and (D).}
    \end{center}
  \label{fig4}
\end{figure}

In both the reduced LJ and WCA systems, the solvent density in the
inter-plate region begins to decrease at the critical separation of the
dewetting transition. For the reduced LJ plate,
the solvent density dramatically decreases when the separation
decreases from 0.95 to 0.92 nm.  Both the variance of water density
and the solvent relaxation time peak at $z_c = 0.92$ nm.
The density of water between the WCA plates
decreases as the separation decreases from $1.45$ to $1.35$ nm. The
ratio of the variance to the average of water density and the
solvent relaxation time peak in the same region.  At the critical
separation for the dewetting transition, the inter-plate region
fluctuates between wet and dry.  The value of the critical distance between surfaces is on the order of one nanometer for the present-sized plates and decreases with decreasing hydrophobicity, in agreement with macroscopic thermodynamic analysis.~\cite{lum97,Margulis:2003wo,sharma12}  This characterization of the density
and the fluctuation of solvent is consistent with the results of
previous studies.~\cite{Margulis:2003wo,joe12}

In the three systems presently studied, it can be clearly seen that
the friction coefficient increases where the solvent fluctuations
become large and slow.  Taken together, both static and dynamic
solvent behavior engender the large frictions at small inter-plate
separations. In agreement with our prior work,~\cite{joe12}
 molecular-scale hydrodynamic interactions
largely result  from such fluctuations when, in the case of
hydrophobic bodies, the drying transition occurs or, for less
hydrophobic species, when water molecules are expelled from the
inter-plate region due to steric repulsion.

\begin{figure}
   \begin{center}
     \includegraphics[scale=0.27]{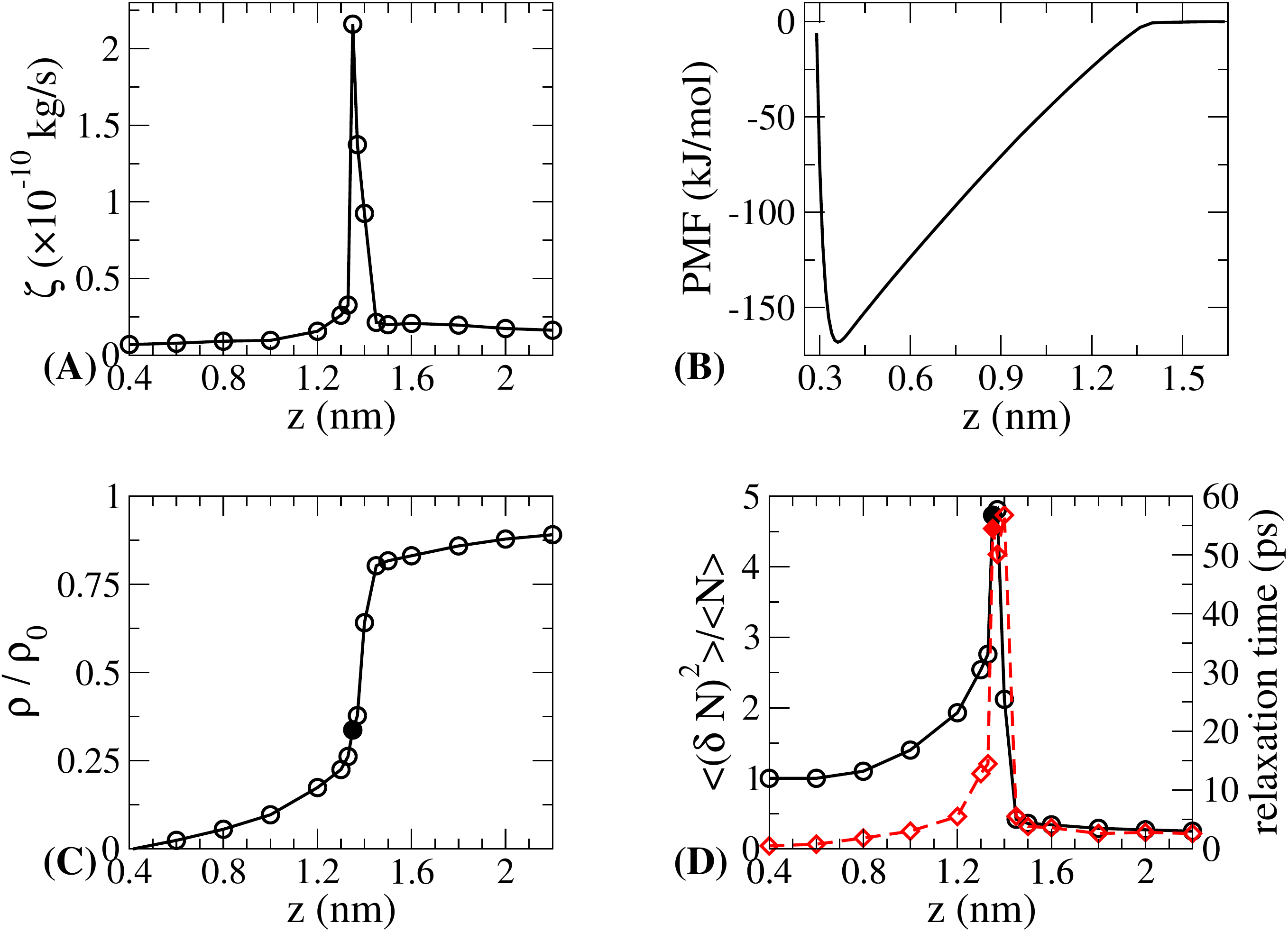}
     \caption{Spatial dependence of relative friction coefficient (A) and
       the potential of mean force (B) for the association of two WCA plates. (C) The relative
       density of water in the inter-plate region. (D) The ratio of
       the variance to the average of the number of water molecules
       (black) and the solvent relaxation time in the
       inter-plate region (red). The value corresponding to the
       friction peak is shown as filled symbols in panels
       (C) and (D).}
    \end{center}
  \label{fig5}
\end{figure}

\subsection{Comparison of Brownian dynamics with molecular dynamics}
\label{sec:association}

In order to further elucidate the impact of molecular-scale
hydrodynamics on the kinetics of self assembly, we perform direct
molecular dynamics simulations of this process. The two plates are
initially placed perpendicular to the z-direction at a separation of
$\sim$1.8 nm. A constant loading force is added to the upper plate and
the lower plate is fixed in its initial position. The loading force
utilized is $440$ pN (full LJ) or $20$ pN (reduced LJ and WCA). There
are at least fifty-five association simulations performed for each
system.

The assembly of both the WCA plates and reduced LJ plates slows down
around the critical separation for dewetting.  This
process is illustrated for the reduced LJ case by the sample
trajectories depicted in Fig. \ref{fig1}.~\cite{vmd96} One can see there is a large
dwell time at $z\approx 1$ nm.  In the case of the full LJ system,
the upper plate initially rapidly approaches the lower plate and the
inter-plate distance decreases to $z=0.95$ nm, corresponding to two
inter-plate water layers.  The association process then stagnates.
After some period a water layer is expelled and
the plate separation decreases to $z=0.66$ nm where
a single water layer separates the plates. There is another dwell
time before the last water layer is expelled and the
two plates finally come into contact. During the association process
the upper plate can rock. This is particularly pronounced around
$z=0.66$ nm where the amplitude of the rocking can be as much as $0.2$ nm.  
In order to provide the best comparison with the
(one-dimensional) BD scheme outlined below, the plates must be kept
as parallel as possible. This is facilitated by the addition of a
harmonic restraining potential on the plate's internal degrees of
freedom. It is important to note that outside the present context,
``rocking'' could be a viable degree of freedom, and we observe that
assembly occurs significantly more rapidly when such effects are
included.  In the case of less attractive or purely repulsive solvent-solute interactions, 
the effect is much less prominent, as the displacement generated by the rocking mode is 
small with  respect to the critical separation for dewetting that drives assembly ($\approx 1$ nm). 

As we will show, the mean first passage times observed in the
MD association trajectories cannot be predicted by solely considering
the free energy profile with constant friction.  In order to evaluate
the role of hydrodynamic interactions in assembly, we utilize a
one-dimensional Brownian dynamics (BD) scheme as described in
Section \ref{sec:HI}, where the system evolves according to Eqn. \ref{eq:BD}
and can be integrated as described by Ermak and
McCammon.~\cite{mccammon78}  The dynamical degree of freedom is taken
to be the inter-plate distance, $z$, along which the friction and
potential of mean force have been computed (see Figs. 3-5).

\begin{table*}[t]
\begin{center}
\begin{tabular}{|c|c|c|c|c|c|c|}
\hline
 type & $F_0$ (pN) & $\bar{\tau}_\text{MD}$ (ps) & $\bar{\tau}_\text{BD-HI}$ (ps) & $\bar{\tau}_\text{BD-NOHI}$ (ps) & $z_0$ (nm) & $z_A$ (nm) \\
\hline
WCA        & 20  & 515  & 516 & 134  & 1.60 & 0.40 \\
\hline
reduced LJ & 20  & 1469 & 2390 & 200 & 1.20 & 0.40\\
\hline
LJ (1st barrier)     & 440 & 573  & 1460 & 85 & 1.20 & 0.80 \\
\hline
LJ (2nd barrier)     & 440 & 32600 & 23078 & 243 & 0.75 & 0.40 \\
\hline
\end{tabular}
\end{center}
\caption{Mean first passage time, $\bar{\tau}$, of the
plate association process as described in Section \ref{sec:association}} \label{tab:mfpt}
\end{table*}

This scheme may be utilized to generate predictions for the mean
first passage time (mfpt) and distribution of first passage times
(dfpt) from BD simulations with the loading force.
Such a comparison serves to evaluate the degree to which the
Brownian framework can yield an adequate description of the kinetics
of assembly.  The constant force $F_0$ applied to the molecular
dynamics simulations is accounted in the BD by a means of the
modified potential of mean force $U(z)$,
\begin{equation}
U(z) = W(z) + F_0z,
\label{eq:modpot}
\end{equation}
where $W(z)$ is the PMF determined from MD in the absence of a
loading force.  Because the loading force $F_0$ in Eqn. \ref{eq:modpot}
is independent of the solution degrees of freedom, Eqn. \ref{eq:modpot}
is rigorous. Moreover, because the loading potential is linear in
$z$ we assume it will not alter the spatial dependence of the
friction. However, for very large loading forces we expect that
non-equilibrium effects will be significant and the
BD picture will fail.  For each system, we have generated 10,000 BD
trajectories given initial state $z_0$, and an absorbing boundary at
$z_A$.  For each system, the values of these parameters and of the
mean first passage time are given in Table \ref{tab:mfpt}.  The
corresponding first passage time distributions are shown in Fig.
6.  The mean first passage time may also be computed directly from
the Smoluchowski equation (Eqn. \ref{eq:smoleq}) by means of the following
expression:~\cite{zwanzig}
\begin{equation}
  \bar{\tau}(z_0) =\beta \int\limits_{z_0}^{z_A} \mathrm{d}y \, \frac{\zeta(y)}{e^{-\beta U(y)}}
  \; \int\limits_{z_R}^{y} \mathrm{d}x \, e^{-\beta U(x)},
\label{eq:mfpt}
\end{equation}
where $\beta\equiv1/k_BT$ and $z_R$ is the positon of the
reflecting boundary.  Due to the the driving
force, the reflecting boundary can be taken to large values in our
calculation.  This equation and an ensemble of BD trajectories will
yield the same result within numerical error.

The mean first passage time obtained from MD simulation for the
association of the reduced LJ plates, is about 60\% smaller than the
result from BD simulation with hydrodynamic interaction (BD with
HI), but 7 times larger than the value estimated from BD when the
spatial dependence of the friction is not included and the single
plate friction (see Table \ref{tab:singlefric}) is utilized instead (BD
without HI).  The fpt distribution obtained from MD is also close
to, albeit narrower than, that obtained from BD
with HI. In contrast, the distribution garnered from BD without HI
is much more strongly peaked.  In the case of the WCA plates, the
mfpt and the fpt distribution from both MD and BD with HI are very
close to each other, but both are approximately 4 times larger than
the BD without HI result.

The results of MD simulation are in reasonable agreement with
those obtained from BD with HI for both the WCA and the reduced LJ plate
systems. The association process slows down in the region around the
critical separation for dewetting transition where the friction
peaks. As discussed above,  the behavior of friction
profile at the critical separation largely results from the solvent
fluctuation due to the dewetting transition.  The molecular-scale
effect of hydrodynamic interaction evidently contributes to the
slowing down of the association process near the critical separation,
and if only barriers present in the PMF are considered (as in BD
without HI), association occurs far too rapidly.  These results
indicate that a kinetic barrier along the reaction coordinate is
present at the drying transition.

\begin{figure}
   \begin{center}
     \includegraphics[scale=0.27]{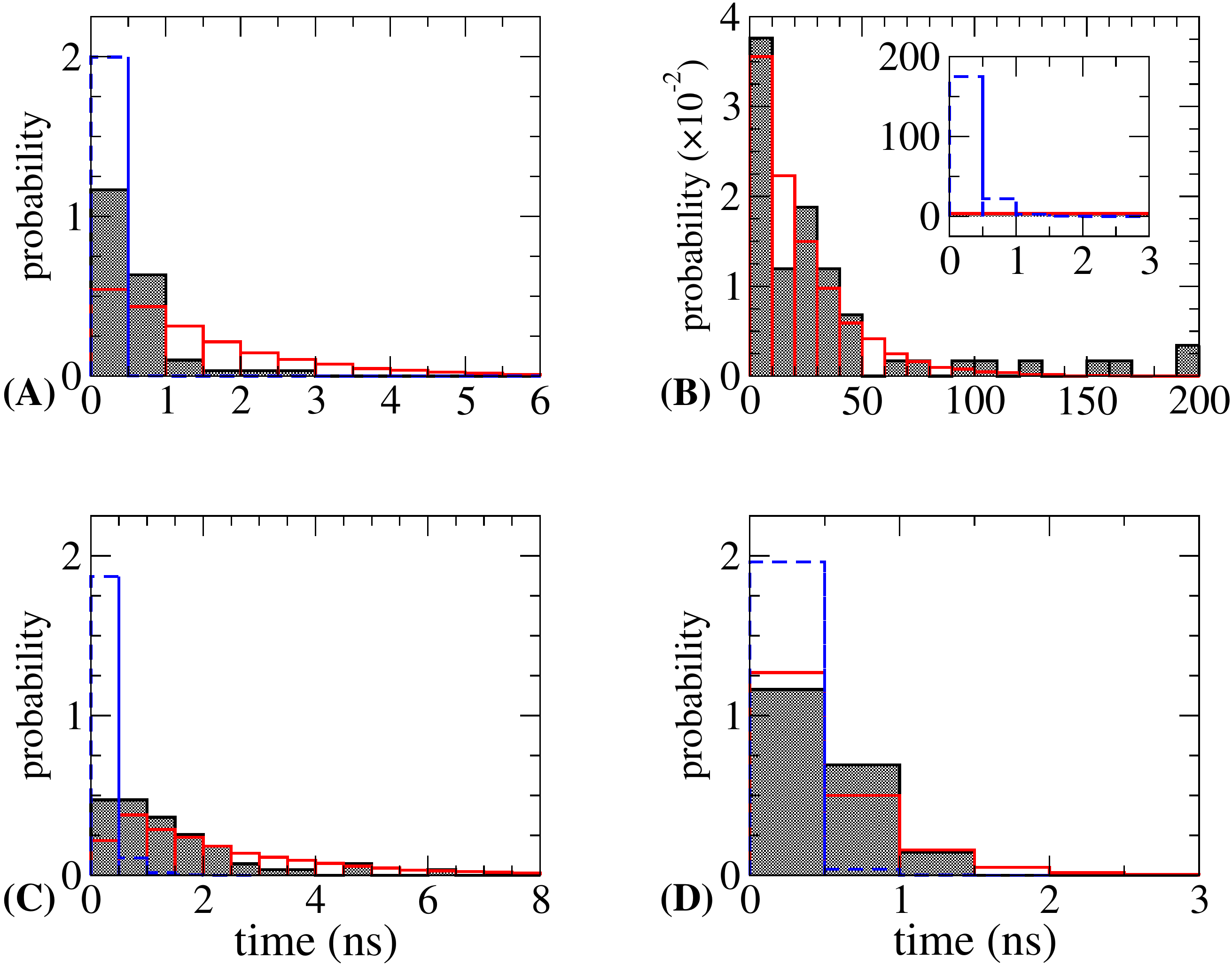}
  \caption{The distribution of first passage time (dfpt) for the association
  process of LJ plates (panel (A), first barrier; panel (B), second barrier), reduced LJ plates (C) and WCA
  plates (D), obtained from the MD simulation (black), BD simulation with and
  without the consideration of hydrodynamic interactions (HI) (red and blue, respectively).
  The dfpts obtained from MD are in reasonable agreement with the result of BD with HI,
  and the probability of having a fpt smaller than 500 ps is much larger in the results garnered BD without HI.}
    \end{center}
  \label{fig6}
\end{figure}

The values of the mfpt of the first (second) barrier of the full LJ
plate system as obtained from BD with HI are 1460 (23078) ps, much
larger than the results of 85 (243) ps for BD without HI. Hydrodynamic interactions strikingly slow down the first
passage time over the second barrier by about two orders of
magnitude. The mfpt obtained from the MD simulation is 573 and 32600
ps, for the first and second barrier, respectively. For the first
barrier, the mfpt obtained from MD simulations
is smaller than the result from BD with HI, while still being much
larger than the result from BD without HI.  Meanwhile, for the second
barrier, the mean first passage time from MD simulations is larger than
the result from BD simulations with HI.  The distribution of the
first passage times corresponding to the second barrier
obtained from MD simulations is similar to the results from BD
simulations with HI. For passage over the first barrier, the distributions exhibit greater deviation.

In general, the average value and the distribution of mean first
passage times of the three types of plates calculated from MD
simulations is consistent with the results of BD simulations with
HI. In prior work,~\cite{joe12} we found that hydrodynamic
interactions contributed approximately 40\% to the assembly of two
fullerenes.  In the present set of calculations, hydrodynamic
interactions contribute a much larger share.  Comparison of spheres
and plates indicate that the contribution of the hydrodynamic
interaction is enhanced as the shape is flattened. More water
molecules are confined by plates than spheres of the same surface area
so that, at small separations, the degree of
confinement and the length scale of
dewetting is increased.

\section{Conclusion}
\label{sec:conclusion}

In this work we study the impact of hydrodynamic effects on the
kinetics of assembly of two plates of varying hydrophobicity.  To
this end, the potential of mean force and spatially dependent
friction coefficient are determined along the inter-plate
separation.  The results show that there is a
correspondence between peaks in the PMF and peaks in the frictional
profile.  High values of the friction are related to large and slow
solvent fluctuations in the inter-plate region. Both solvent
confinement and drying phenomena can play a critical role in the
kinetics of assembly. In this way, molecular scale effects shape the
hydrodynamic interactions, and give rise to
their deviation from continuum theory,  which predicts 
that the friction coefficient diverges as two 
bodies come into contact.~\cite{wolynes76,Jeffrey:1984tu}

The kinetics of assembly studied by means of molecular dynamics
simulations in explicit solvent with a constant loading force applied
along the reaction coordinate is compared with the predictions of
Brownian dynamics with the same loading force and with the
potential of mean force and hydrodynamic profile extracted from the data presented in
Section \ref{sec:fricresults}.
Brownian dynamics is a widely used technique owing to its
computational efficiency as solvent degrees of freedom are not treated
explicitly. There is reasonable agreement between the mean first
passage times that are obtained from the two schemes. Indeed, we find
that hydrodynamic interactions are essential to produce a reasonable
description of the process and neglect of spatial dependence of the
friction has a large impact on the kinetics.  The HI give rise kinetic
bottlenecks along the association pathway,~\cite{joe12} over and above
the barriers in the potential of mean force. Interestingly, other
degrees of freedom such as plate ``rocking'' can in some cases significantly
increase the rate of assembly, probably by facilitating waters entry
and exit from the inter-plate gap.  In order to describe the effect of solvent relaxation on the rocking mode within the Brownian framework, it would require a determination of the friction coefficient experienced on this degree of freedom.  The exploration of all possible pathways to plate assembly is beyond the scope of this work, and
presently we concentrate on the approach of two parallel plates.

It has been suggested~\cite{Margulis:2003wo,willard08} that
including solvent degrees of freedom in the reaction coordinate is
necessary in order to properly characterize the association process.
Presently, a reaction coordinate is utilized that is only a function
of the plate degrees of freedom.  We have shown that such a choice
is reasonable provided that (molecular-scale) hydrodynamic
interactions are considered along the pathway.  The inclusion of
hydrodynamic effects captures the effect of solvent in an average
sense, and the Brownian framework would be an exact treatment in the
limit where the solvent time scales are much faster than those
associated with the heavy bodies. Leading from the last point, the
discrepancies between the MD and BD results can be largely
attributed to either the impact of the pulling force, that is the
friction experienced along the force-biased surface significantly differs from that extracted from the
equilibrium calculations, or to non-Markovian effects. In the case
of the passage over the first barrier of the LJ system, at least
some of the deviation is likely due to the large pulling force,
although non-Markovian effects should have an impact both in this
case and in the other system studied.  It has, for
instance, been shown that dynamical caging effects can be exhibited
along degrees of freedom that are associated with slowly varying
memory functions.~\cite{tuckerman} Such phenomena can also serve to explain
the long dwell times exhibited in the association trajectories.

Unfortunately, given the broad distributions involved and limited
number of MD trajectories that can be harvested, it is difficult to
fully gauge the impact of non-Markovian effects based on the present
work.  Certainly, it is observed that, in the case of strongly
hydrophobic plates, assembly is almost always preceded by the
``first'' drying transition, that is the system does not sample
successive wet and dry states near the critical separation.  One
should expect such a process to be intrinsically non-Markovian,
although the BD with HI approach that is presently employed appears
to at least capture some aspect of this ``waiting period'' for
drying in an average way.  A more precise understanding of this
phenomena will be the subject of future work.

\begin{acknowledgements} This research was supported by a grant to
  B.J.B. from the National Science Foundation via Grant No. NSF-CHE-
  0910943. J.L. was supported by the CAS Hundred Elite Program and MOST 973 program No.2012CB932504.
\end{acknowledgements}

\appendix
\section{Continuum hydrodynamic treatment of the friction on a rigid
body}
\label{sec:appendix}

Consider a rigid body made up of $N$ spherical sites where the origin is taken to be the
center of mass.  A $3N \times
3N$ mobility tensor $\mat{B}$ may be defined where:
\begin{equation} {\bf V}= \mat{B}\bullet{\bf F},
\end{equation}
and where $\bf V$ and $\bf F$ are $3N$ dimensional vectors
representing the velocities and external forces on the $N$ sites that comprise the body.
In the case of stick boundary conditions, the elements of $\mat{B}$ are approximated by the
Rotne-Prager tensor~\cite{rotne69} where the $3\times 3$ elements, $\mat{B}_{ij}$, are given as:
\begin{equation} \mat{B}_{ij} = \frac{1}{6\pi\eta a} \left[
\delta_{ij} + (1-\delta_{ij}) \mat{T}_{ij} \right], \label{eq:bij}
\end{equation}
where $a$ is the site radius, and $\eta$ is the shear viscosity. For TIP4P water the value is
$\eta = 0.494$ $\mathrm{mPa}\,\mathrm{s}$ and has been taken from the literature.~\cite{viscoref}
The matrix $\mat{T}_{ij}$ is given by the following expression:
\begin{widetext}\begin{equation}
\mat{T}_{ij} = \Bigg\{ \begin{array}{cc}
\frac{3}{4}\frac{a}{r_{ij}}\left( \mat{I} + \mat{\hat{r}\hat{r}}_{ij}\right) +
\frac{1}{2} \left(\frac{a}{r_{ij}}\right)^3 \left( \mat{I} - 3\,
\mat{\hat{r}\hat{r}}_{ij} \right) & r_{ij} > 2a \\ \left(1 - \frac{9}{32}
\frac{r_{ij}}{a}\right)\mat{I} + \left( \frac{3}{32} \frac{r_{ij}}{a} \right) \;
\mat{\hat{r}\hat{r}}_{ij} & r_{ij} \leq 2a \end{array},
\end{equation} \end{widetext}
where  $r_{ij} = \left| \vec{r}_{ij} \right|$ and $\mat{\hat{r}\hat{r}}_{ij}$ is the vector direct product of the unit vector of displacement.
If the plate is treated as a body that is centro- and
axi- symmetric, then the translational friction tensor, $\mat{\zeta}_{T}$, is given by the following expression:~\cite{carrasco99}
\begin{equation} \mat{\zeta}_{T} = \sum\limits_{ij} \left[
\mat{B}^{-1}\right]_{ij},
\end{equation}
where $\mat{\zeta}_{T}$ is a diagonal $3\times 3$ matrix.  Presently, we are
interested in the friction coefficient associated with motion perpendicular to
the face of the plates, and this is what is reported in
Section \ref{sec:singleplate}.
For periodic systems, the elements of $\mat{B}$ are given by
\begin{equation} \mat{B}_{ij}^\text{pbc} = \sum\limits_{\vec{n}}
\; \mat{B}_{ij}^{\vec{n}}\; ,
\end{equation}
where $\vec{n}$ is the periodic image vector, and $B_{ij}^{\vec{n}}$ the same as in Eqn. \ref{eq:bij} except it is evaluated over periodic images.  This expression may be evaluated by means of an Ewald
summation, as was shown by Beenakker.~\cite{Beenakker:1986tz}  In the case of overlapping
spheres, reciprocal space contributions are excluded from the Ewald
sum, as the long-range portion of the hydrodynamic interaction only includes terms for which $r_{ij}>2a$.

\providecommand*\mcitethebibliography{\thebibliography}
\csname @ifundefined\endcsname{endmcitethebibliography}
  {\let\endmcitethebibliography\endthebibliography}{}


\end{document}